\documentclass[journal=jctcce,manuscript=article,layout=traditional]{achemso}
\pdfoutput=1
\usepackage{graphicx}
\usepackage{url}
\usepackage{hyperref}

\newcommand{\A}{\mbox{ \AA}}
\newcommand{\ksmd}{\mbox{K}^+_\mathrm{SMD}}

\newcommand{\REF}{Ref. \citep{DeFabritiis2008}}

\author{Toni Giorgino}
\email{toni.giorgino@upf.edu} 
\affiliation{Computational
  Biochemistry and Biophysics Laboratory (GRIB-IMIM), Universitat Pompeu
  Fabra, Barcelona Biomedical Research Park (PRBB), C/ Doctor Aiguader
  88, 08003 Barcelona, Spain}
\author{Gianni De Fabritiis}
\email{gianni.defabritiis@upf.edu} 
\affiliation{Computational
  Biochemistry and Biophysics Laboratory (GRIB-IMIM), Universitat Pompeu
  Fabra, Barcelona Biomedical Research Park (PRBB), C/ Doctor Aiguader
  88, 08003 Barcelona, Spain}

\title{A high-throughput steered molecular dynamics study on the free
  energy profile of ion permeation through gramicidin A}

\begin{document}

\begin{abstract}
Steered molecular dynamics (SMD) simulations for the calculation of
free energies are well suited for high-throughput molecular
simulations on a distributed infrastructure due to the simplicity of
the setup and parallel granularity of the runs. However, so far, the
computational cost limited the estimation of the free energy typically
over just few tens of pullings, thus impeding the evaluation of
statistical uncertainties involved.  In this work, we performed two
thousands pulls for the permeation of a potassium ion in the
gramicidin A pore by all-atom molecular dynamics in order to assess
the bidirectional SMD protocol with a proper amount of sampling.  The
estimated free energy profile still shows a statistical error of
several kcal/mol while the work distributions are estimated to be
non-Gaussian at pulling speeds of 10 \AA/ns. We discuss the
methodology and the confidence intervals in relation to increasing
amounts of computed trajectories and how different permeation pathways
for the potassium ion, knock-on and sideways, affect the sampling and
the free energy estimation.
\end{abstract}
\maketitle

\section{Introduction}

Biologically relevant events often take place at time scales far beyond those
accessible by fully atomistic simulations, for
example,  conduction  of ions through narrow
channels~\cite{Allen2003,Jordan2005,Allen2006a}.
A successful approach for describing molecular phenomena at longer timescales
is to average out all but a few degrees of freedom of the system by selecting a
{\em reaction coordinate}.  The forces affecting the
process are then described as an effective potential of mean force 
(PMF), i.e.\  the free energy profile along the reaction
coordinate~\cite{Aqvist2000,Treptow2006,Aqvist1989}.  The PMF can, in principle,
be computed by sampling the equilibrium statistical distribution of the system.
However, the time required for the system to cross
high free energy barriers may be long enough to make the computation
infeasible. 

This problem has been successfully addressed with biasing 
protocols\cite{gervasio_flexible_2005,zwier_reaching_2010},
such as umbrella sampling (US)~\cite{Torrie1977}, which overcomes this
limitation by sampling several biased {\em equilibrium} distributions,
which are later merged by histogram-based
techniques~\cite{Berneche2001,Roux1995,Kumar1992,Kato2005}.  Jarzynski
equation (JE) and Crooks fluctuation theorem (CFT) equalities showed
that the PMF can also be recovered from {\em non-equilibrium} steered
molecular dynamics~\cite{Crooks2000,Jarzynski1997}.  Steered molecular
dynamics (SMD) is a well-known computational protocol to exploit
non-equilibrium sampling, in which the application of time-dependent
biasing forces guides the system according to a predefined
protocol~\cite{Izrailev1998,Isralewitz2001}.  In SMD experiments,
several pulls are simulated in one
(forward)~\cite{Jarzynski1997,Hummer2001} or two (forward and reverse)
directions~\cite{Forney2008,Kosztin2006,Crooks2000,Minh2008}.

A number of previous studies have used SMD
simulations to compute free energy profiles on realistic biomolecules.
The JE applied to {\em one-directional} (forward only) SMD experiments
have been used by several authors to compute the free energy profile
in large biomolecular systems: among the most recent works, Cuendet et
al.~\cite{Cuendet2008} used two groups of $n \sim 150$
single-directional trajectories (total sampled time of approximately 2
$\mu$s) to compute the PMF of the T cell receptor with a major
histocompatibility complex peptide   (TCR-pMHC) complex and a
mutant. Martin et al.~\cite{Martin2009} used single-directional SMD on
a large system to compute the energetics of translocation of a
polynucleotide through a nanopore; the pullings were conducted with
various parameters, and the PMF curves computed on 2 to 6 samples.
Liu et al.~\cite{Liu2006} studied the permeation of Na$^{+}$ through gramicidin A
(gA) with $n \simeq 8$ single-directional trajectories; Zhang et
al.~\cite{Zhang2006} used $n \sim 35$ trajectories and four different
computational methods based on the JE to compute the unbinding of
acetylcholine from the Alpha-7 nicotinic receptor along four different
paths.  Jensen et al.\ computed the energetics of sugar permeation
through lactose permease~\cite{Jensen2007} and glycerol through
aquaglyceroporin~\cite{Jensen2002}, respectively, via four SMD runs
with cumulant expansion.
Comparatively, fewer works
have discussed the use of {\em bidirectional} pulling experiments in
large proteins.  De Fabritiis et
al.~\cite{DeFabritiis2008} computed the PMF in gA with the
with 25 pulls in each direction. Forney et al.~\cite{Forney2008} 
 computed the PMF of gA, as well, with 10 pulls in each direction
comparing various ionic strengths and backbone restrain types.
Due to limitations in computational resources, most studies could only
sample a  low number of trajectories, and were therefore
unable to assess the impact of increased sampling on the precision of
the PMF profiles.

Here, we expand the amount of sampling with respect
to previous bidirectional SMD studies by almost two orders of
magnitude in a realistic system, the gA dimer embedded in a
membrane and explicit solvent, to test the methodology in a properly
sampled system. Gramicidin A (see \ref{fig:ga-axes}) is a helical antibacterial dimer
(15 aminoacids each) which increases permeability of biological
membranes to inorganic ions~\cite{miloshevsky_gating_2004}.  The
backbone of the gA dimer form a narrow pore, allowing a single file of
water molecules (or potassium ions) to fill it.  Due to the diameter
of the pore, the transport of a single ion drags with it a column of
six to nine water molecules in a single file, reducing drastically the
possibility that water molecules slip past each other~\cite{hille01}.
The small size and early availability of its structure made gA a
frequently-used model for a membrane channel~\cite{Chung02,Mamonov03}.
Despite its simplicity, permeation is not so well reproduced
computationally: several studies reported with a barrier to permeation
of 10-20 kcal/mol, several kcal/mol higher than the experimental
one~\cite{degroot02,Allen2006a,Bastug2007}. The barrier height was
recently shown to be much improved with the use of a polarizable
forcefield~\cite{patel_exploring_2009}.

 We performed an extensive set of all-atom MD
experiments on the gA channel~\cite{Andersen2005} and
computed the PMF from bidirectional pulling
experiments~\cite{Minh2008,Ytreberg2006} using 1,000 pulls per
direction. Confidence bands for increased number of pulls, computed
with a variable-size bootstrap procedure, are also presented. The
importance of sampling effectively the degrees of freedom orthogonal
to the reaction coordinate is well illustrated by two permeation
pathways shown by the potassium ion in the interface between the two
monomers of gA with different free energy profiles.

\section{Materials and methods}

\subsection{The potential of mean force}

The PMF is a convenient description of the energetics of the system
obtained integrating out all of the degrees of freedom with the
exception of one {\em reaction coordinate}, $z=z({\bf R})$, which
should capture the interesting features of the system. The
PMF $G(z)$ would then be
$$
e^{-\beta \Delta G(z')} = \frac{\int d{\bf R}d{\bf P} \delta(z({\bf R})-z')
  \exp(-\beta H)} {\int d{\bf R}d{\bf P} \exp(-\beta H)},
$$
where $H=H({\bf R},{\bf P})$ is the Hamiltonian of the system, ${\bf
  R}=(\mathbf{r}_1,...,\mathbf{r}_N)$, ${\bf
  P}=(\mathbf{p}_1,...,\mathbf{p}_N)$ are the  positions and
momenta of the $N$  atoms, $\delta(\cdot)$ is the
Dirac delta function and $\beta = 1/(k_B T)$, where $k_B$ is the Boltzmann
constant and $T$ the temperature of the system.

The Crooks fluctuation relation~\cite{Crooks1999,Hummer2001,Collin2005}
allows one to compute the equilibrium free energy difference $\Delta
G$ between two states ``0'' and ``1'' described by two Hamiltonians
$H_0$ and $H_1$ as
\begin{equation}
  \frac{P_F(+\beta W)}{P_R(-\beta W)} = \exp(\beta ( W - \Delta G)),
  \label{eq:crooks}
\end{equation}
where $W$ is the external work done on the system by forcing it to change from
state $0$ to $1$, and $P_F$,$P_R$ are the probability distributions of releasing
the work $W$ into the system during a transformation in the forward (F) $0
\rightarrow 1$ and reverse (R) $1\rightarrow 0$ direction, respectively, in a
finite time.  The Crooks fluctuation relation is a generalization of  the Jarzynski equality
(JE)~\cite{Jarzynski1997}, 
$$
  \left< \exp(-\beta W) \right> _F=\exp(-\beta \Delta G),
$$
recovered from Crooks fluctuation relation by integrating both sides
of Eq.~\ref{eq:crooks}. The Crooks fluctuation relation can be estimated using  the optimal
Bennett acceptance ratio method~\cite{Bennett1976}. 
Interestingly, these two fundamental relations have their equilibrium
counterparts obtained for an infinite pulling speed, where CFT resembles  an
equilibrium relation previously derived by Shing and Gubbins~\cite{Shing82} and
JE corresponds to Widom's formula used to compute the chemical potential by
test particle insertion~\cite{delgado-buscalioni_determination_2005} (with  well known
poor convergence properties~\cite{FrenkelBOOK}). The exponential average
of the JE causes few rare low-energy trajectories dominate
the estimate of $\Delta F$; when only few trajectories are available,
the estimate can be improved using a first-order cumulant expansion,
which is exact in the limit when the distribution of $W$ values is
Gaussian~\cite{Gore2003,Cuendet2008}. The CFT have much better 
convergence properties than a direct application of JE, and was therefore used 
in a previous study~\cite{DeFabritiis2008} and in this study. Both the JE
and the CFT  have
been confirmed experimentally in atomic force and
single-molecule pulling experiments~\cite{Hummer2001,Collin2005}.

In this work, we use the estimator for the PMF $G(z)$ proposed by Minh
and Adib~\cite{Minh2008}, 
\begin{eqnarray} \label{eq:Gz}
&& e^{-\beta G(z)} =   \nonumber \\
&& \qquad = \sum_t 
  \left[
     \left< \frac{ n_F \, \delta(z-z_t) e^{-\beta W_0^t}}{n_F + n_R
       e^{-\beta(W-\Delta F)}} \right>_F + \right. \nonumber \\
&& \qquad \qquad  +  \left. \left< \frac{ n_R \, \delta(z-z_{\tau-t}) e^{\beta W_{\tau - t}^\tau}} {n_F + n_R e^{\beta(W+\Delta F)}} \right>_R
  \right] \times \nonumber \\
&& \qquad \qquad \times \, e^{\beta \Delta F_t} 
  \Bigg/ \sum_t e^{-\beta [V(z;t) - \Delta F_t]}   
\end{eqnarray}
where $n_F$ and $n_R$ are the number of forward and reverse
trajectories respectively, $W_a^b$ is the partial work performed in
the interval between time $a$ and $b$, $\tau$ is the final simulation
time, and $\Delta F_t$ is the free energy difference between the
initial state and the one at time $t$. 
In this work we used the implementation of  Eq.~(\ref{eq:Gz}) provided 
in the FERBE  package~\cite{MinhFerbe}.

\subsection{Steered MD Protocol}

To describe the permeation experiments of one potassium ion through
the gA channel, the collective reaction coordinate was assumed to be
the $z$ coordinate of one of the cations, which will be called $\ksmd$
in the following, i.e.\ $z({\bf R}) = z_{K}$. The chosen cation was
driven through the channel  applying a
simple harmonic biasing potential parallel to the $z$ axis producing the force
\begin{equation}
  F_z(z_K,t)= -k \big( z_K - b(t) \big) \label{eq:force}
\end{equation}
where $k$ is the spring constant, $z_K$ is the instantaneous $z$
coordinate of the ion, and $b(t)$ is the time-dependent equilibrium
point for the biasing force.  
The spring constant $k=10$ kcal/mol of the biased system was set in order to
fulfill the strong spring approximation, $k>\max(\frac{2\alpha}{\delta
  z^2},\frac{2U_{max}}{\delta z^2})$, where $\delta z$ is the spatial
resolution that we are seeking for the PMF, $U_{max}$ is the maximum
energetic barrier that we expect in $\delta z$, and $\alpha \gg 1$.
The biasing position was displaced
linearly with time:
$$
b(t) = \left\{
    \begin{array}{lc}
      z_F + v t & \mbox{(forward)} \\
      z_R - v t & \mbox{(reverse)}
    \end{array}
    \right.
$$

The cumulative work profile was
obtained by integrating the instantaneous forces over the corresponding
interval
\begin{equation}
  W(t)=\int_{t'=0}^T F(t') \, v \, dt'  \label{eq:W}
\end{equation}
with $F(t)$ given by Eq.~(\ref{eq:force}).  For the numerical
computation of Eq.~(\ref{eq:W}), the integral was approximated as a
discrete summation over time intervals of length $\Delta t$,
\begin{equation}
  W_a^b =  \sum_{t_j=a}^{b}  -k \big(
    z(t_j)-z_0 \mp v t_j \big)  \, v \, \Delta t
    \label{eq:Wdiscrete}
\end{equation}
where $t_j= j \, \Delta t $ is the time corresponding to the $j$-th
interval, $z_0$ is the starting position of the pull, and the sign is
taken according to the pull direction.  The $z$ axis was divided in
100 bins, equally spaced over the interval $z=-10, \, \dots , 0
\A$.

\subsection{Preparation of the system}

\begin{figure}[tb]
\begin{center}
  \includegraphics*[width=.8\columnwidth]{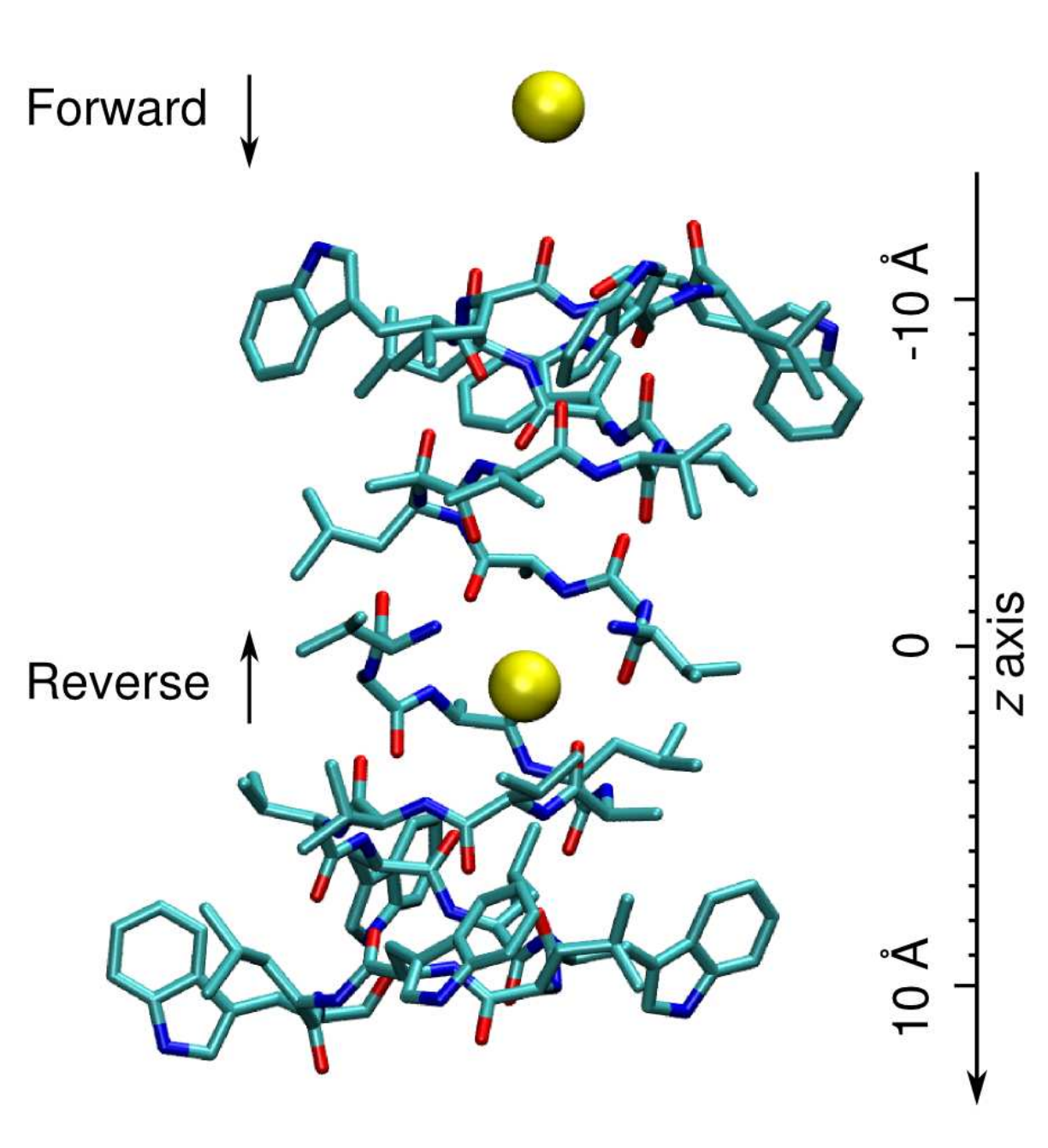}
\caption{ Structure of the gramicidin A dimer, with the starting K$^+$
  ion positions for the forward and reverse steered molecular dynamics
  runs (yellow). In the former, the cation is pushed from the outside
  towards the inside of the channel ($z$ increasing from $-15$
  \AA\ towards 0 \AA). In the reverse simulations, the ion is pulled
  along the time-reversed path. The $z$ axis runs through the center
  of the pore. Lipid bilayer (117 DMPC molecules), water (8,668
  molecules), and other solvated ions have been omitted for
  clarity.  }
\label{fig:ga-axes}
\end{center}
\end{figure}

The gA dimer was prepared based on the Protein Data Bank entry
PDB:1JNO\cite{townsley_structures_2001}, 
extending the protocol already presented in De Fabritiis et
al.~\cite{DeFabritiis2008}. The structure used in the previous study,
comprising the gA dimer and dimyristoylphosphatidylcholine (DMPC)
lipid bilayer, was solvated with 8,668 TIP3 water molecules and ionized
at a ionic strength of 150 mM with 24 pairs of K$^+$ and Cl$^-$
ions. The final system, comprising 40,410 atoms, 
was then equilibrated at 1 atm and 305~K with the
CHARMM27~\cite{MacKerell2000} force field in the NPT ensemble for
approximately 13 ns.  The lipid bilayer is oriented in the $xy$ plane,
and the $z$ axis goes through the gA pore (see
\ref{fig:ga-axes}).  The simulation box resulting from the NPT
equilibration was 66.1 $\times$ 65.8 $\times$ 88.9 \AA$^3$.  The
preparation runs were performed with the NAMD
program~\cite{Phillips2005}, with  particle-mesh Ewald electrostatics~\cite{Darden1993},
rigid bonds, cubic periodic boundary conditions, and a time step of 2
fs.
 
In order to generate the initial configurations for the forward run,
the position of one potassium ion was exchanged with that of a water
molecule located close to the entrance of the pore, i.e. approximately
at $(0,0,-15)$ \AA. For the reverse runs, the ion was exchanged with
the water molecule closest to the middle of the channel. The two
systems were subject to a further 20 ns of equilibration in the NVT
ensemble, while restraining $\ksmd$ to its initial position with a
spring constant of 10 kcal/mol/\AA$^2$. After the initial 20 ns of
equilibration, the runs were extended for further 20 ns in the same
conditions, taking snapshots at 200 ps intervals, thus yielding 100
snapshots for each of the two systems. Each snapshot was used as an
initial configuration for 10 SMD runs.  \ref{fig:ga-axes} shows
the ion at the initial positions $z_F$ and $z_R$ for one of the
forward and reverse pulls respectively.

Further analysis on the configurations, e.g.\ the water
occupancy of the pore, and statistics on ion-water-protein relative
positions, were performed using the scripting facilities of the VMD
program~\cite{Humphrey1996}.

\subsection{Production runs}

We performed 1000 forward and 1000 reverse SMD runs, starting from the
200 distinct initial configurations prepared according to the protocol
explained above. The pulling speed and SMD spring constants were set
at $v=\pm 10$ \AA/ns and $k= 10$~kcal/mol/\AA$^2$ respectively.  
Each SMD run lasted 2 ns, long enough for the pulled ion to reach the
starting position for the opposite direction, i.e.\ until $z<z_R$ for
the forward runs, and $z>z_F$ for the reverse ones. When the ion
was outside of the channel, a flat bottom potential was applied to
keep the ion in line with the pore during the approach (the region
$z<10$ was excluded from the PMF calculation).
No constraints were imposed to the ion on the plane orthogonal to $z$.
Only the {\em center of mass} of the C$\alpha$ atoms of the pore was
restrained to its initial position with a harmonic potential of 100
kcal/mol/\AA$^2$, to avoid the pore being displaced out of the
membrane by the net force applied throughout the simulations.  The
restrain was just applied to the center of mass, in order not to
artificially constrain the local helix radius, inter-dimer distance, nor
orientations of the side-chain. The pore was therefore free to
expand under the influence of the permeating cation; this flexibility
has been show to play an important role in the permeation
energetics~\cite{bastug_role_2006,corry_influence_2005,Forney2008}.

All of the production runs were performed with  
ACEMD~\cite{Harvey2009}, which leverages off-the-shelf accelerated
graphic processing units (GPUs) allowing one to achieve approximately
100 ns per day of simulated time on a single GPU for system size of the
order of 23,000 atoms, performance decreasing linearly with system
size.  Production runs have been performed in the $NVT$ ensemble,
Langevin thermostat at 305~K with a relaxation of 0.1/ps, computing
the electrostatic interactions with the particle-mesh Ewalds (PME)
algorithm~\cite{harvey_implementation_2009}.  The integration timestep
was set to 4 fs, enabled by the hydrogen mass repartitioning
scheme~\cite{feenstra_improving_1999,Harvey2009} available in
ACEMD. This scheme allows for longer timesteps by using the property
that the equilibrium distribution is not affected by individual atom
masses provided that the total mass of the system stays the
same. Transport properties change  by less than 10\%, a small
amount compared of the errors  intrinsic in the TIP3P water
model compared to real water\cite{feenstra_improving_1999}.

The runs were performed on a distributed computing grid called
GPUGRID.net~\cite{buch_high-throughput_2010}. We set up a server based
on the Berkeley Open Infrastructure for Network
Computing (BOINC) to automatically distribute the
runs through the Internet~\cite{giorgino_distributed_2010}.   In order to be
executed remotely, each forward and reverse run was arranged as a
separate work unit. As soon as each participating computer finished
computing the assigned simulation time, it returned a log file with
the trajectory $z(t)$, and the force
$F(t)$ exerted on the SMD ion, recorded at intervals of 200 fs.  The
 computational effort used for computing the PMF curves amounted to 
the generation of 4 $\mu$s
of total simulation data.

\subsection{Bootstrapping procedure} \label{sec:bootstr-proc}
The convergence properties of the PMF were estimated with respect to
increased configuration sampling by re-computing the potential curves
with a varying-sample size bootstrapping technique~\cite{efron_bootstrap_1986},
similar to the one employed by Cuendet et al.~\cite{Cuendet2008}.  In
this procedure we constructed resampled sets of the available
bidirectional pulls
of increasing cardinality. Each of the available pulling trajectories
was randomly taken zero or more times, in order to build a resampled
set containing $R$ bidirectional trajectories.  A PMF profile was
computed considering only the resampled set, and the PMF depth
was obtained.  The process was iterated until $B$
bootstrapped replicas were obtained, finally yielding the standard
deviation of the PMF depth, $\sigma(R)$.  The procedure was repeated
for resampled sets of sizes $R=10,50,100,250,500,750,n$, with $n=1000$
being the count of all available pulls. The case $R=1000$ corresponds to
the plain bootstrap procedure, which creates resampled sets as large
as the number of trajectories originally available.

\section{Results}

\begin{figure}[tb]
\begin{center}
  \includegraphics[width=.9\columnwidth]{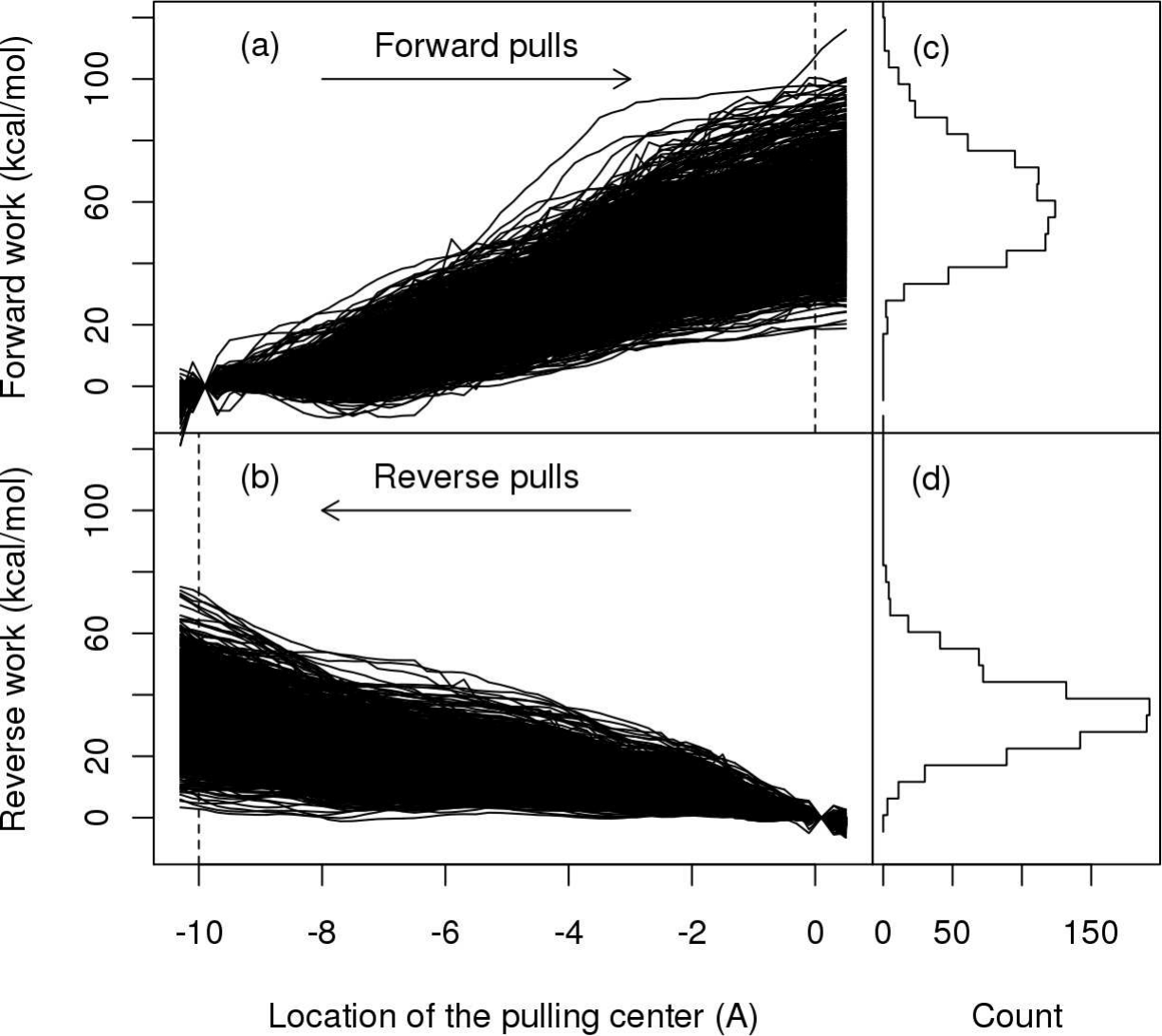}
  \caption{ Profiles of the accumulated work $W(z)$ spent to pull the
    ion inside (a, Forward) or outside (b, Reverse) the gA channel
    at~10 \AA/ns (1000 pulls per direction). The panels on the
    right-hand side show the distribution of final work values for the
    forward (c) and reverse (d) directions respectively. The mean work
    performed in the forward direction at the end of the pulls was
    55.5 kcal/mol (standard deviation 14.9 kcal/mol); for the reverse
    direction it was 31.9 kcal/mol (SD 11.2 kcal/mol).  }
\label{fig:wprofile10}
\end{center}
\end{figure}

We computed cumulative work profiles  using
Eq.~(\ref{eq:Wdiscrete}) for all of the 2000 pulling experiments. The
upper and lower part of \ref{fig:wprofile10} show the work profiles
for the forward (a) and reverse work (b) respectively. For the forward
pulls, the work was taken as the one required to push the $\ksmd$ ion
from $z=-10$ to $z=0$; for the reverse pulls, the endpoints are
reversed.  The distributions of final work values for the forward and
reverse pulls are shown in the right-hand side of
\ref{fig:wprofile10}, respectively in panels (c) and (d). The mean
work performed in the forward direction at the end of the pulls
(dashed line) was 55.5 kcal/mol (standard deviation 14.9 kcal/mol);
for the reverse direction it was 31.9 kcal/mol (SD 11.2 kcal/mol).

\subsection{Final work distributions}

The CFT implies that if the forward work values follow a normal
distribution with a variance $\sigma$, the reverse work values should
also be a Gaussian with the same variance\cite{procacci_crooks_2006}.  To check whether the final
work values obtained in the simulations follow a normal distribution,
we applied the well-known Shapiro-Wilk normality
test~\cite{shapiro_analysis_1965}. The test quantifies the probability
$p$ that a given set of values could have been taken from a Gaussian
distribution (null hypothesis); if the $p$ value computed by the test
is smaller than a fixed threshold, usually taken as 0.05, one
concludes that there is strong evidence against normality.  The
Shapiro-Wilk test rejected the null hypothesis that the final work
values follow normal distributions in either direction ($p < 4 \cdot
10^{-7}$ for both the forward and reverse runs). Given that the  pulling
speed used  here was 10 {\AA}/ns, it is  unlikely that
the work distribution is Gaussian, at least for a system with enough
dissipation like pulling an ion through a pore.

\subsection{PMF profiles}

\begin{figure}[tb]
\begin{center}
 \includegraphics[width=.9\columnwidth]{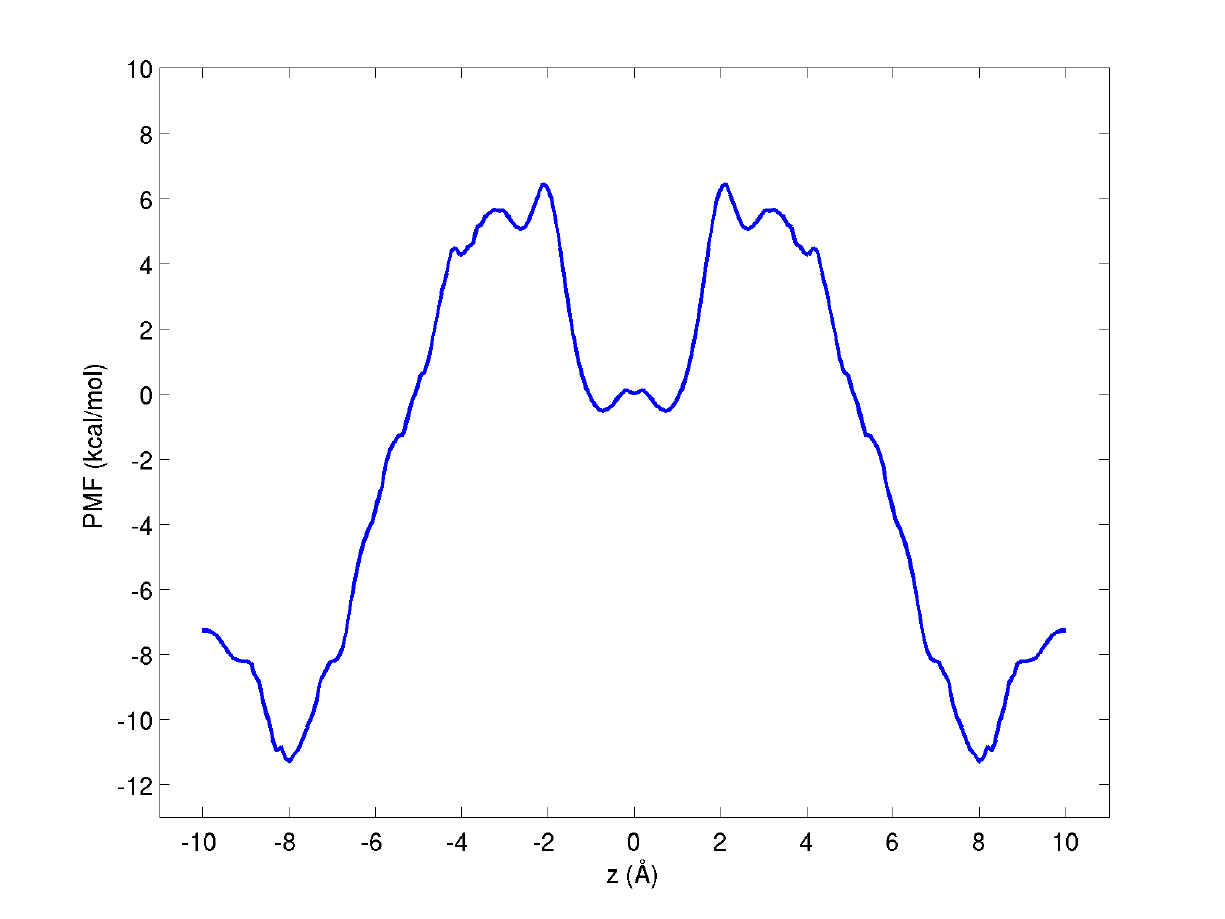}
\caption{Potential of mean force (PMF) curves for a K$^+$ atom to
  cross the gA channel, computed from 1,000 steered MD experiments at
  pulling speed of 10 \AA/ns in each direction.}
\label{fig:pmf}
\end{center}
\end{figure}

The PMF curves recovered from the
bidirectional pulls with the analysis protocol cited
above~\cite{Minh2008} are shown in \ref{fig:pmf}.  
For clarity, the potential profile has been
symmetrized around the $z=0$ axis, and offset so that the PMF is zero
at $z=-10 \A$.  The PMF exhibits a binding site at
$z \simeq 8.5 \A$ and a total barrier height with respect to the bulk of
approximately 14 kcal/mol.  The location of the binding site is
approximately consistent with the ones reported in the
literature~\cite{Allen2004,Bastug2007,Forney2008,patel_exploring_2009}.  

The barrier height obtained here is lower with respect to the one
computed from the \REF\ data set (19 kcal/mol), obtained from 25
bidirectional pulls. As discussed in the same paper, the induced
dipole of the water molecules surrounding the channel provides an
important component to the barrier to permeation. In particular, when
the K$^+$ ion is not in the middle of the channel, a large fraction of
its electrostatic interaction energy is due to atoms between 6 and 16
\AA\ of distance, i.e. in the second coordination shell. This fact
underlines the role played by the water molecules' polarization and
the finite time required for their reorientation.  Part of the
relatively higher barrier to permeation found in this study with
respect to others performed with US~\cite{Allen2004,Bastug2007} or
SMD~\cite{Forney2008} may be therefore ascribed to the biological (150
mM) KCl ionic strength employed outside the pore,
compared to higher concentrations used in other studies.

Furthermore, the previous study \REF\ sampled only 25 bidirectional
pulls, and therefore could not provide a measure of the statistical
uncertainty. We shall show later that that this amount of sampling
still incurs in a statistical uncertainty of  tens of kcal/mol.

\subsection{Permeation pathways}

\begin{figure}
\begin{center}
 \includegraphics[width=.9\columnwidth]{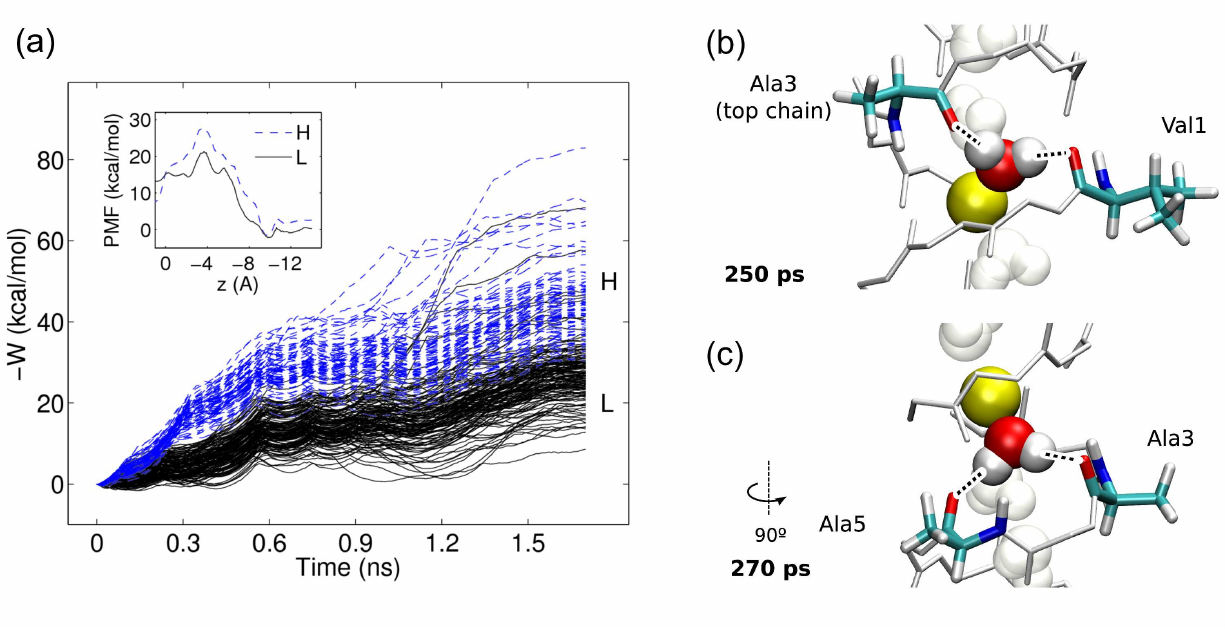}
 \caption{(a) Work profiles for the pulls in groups H (37\% of the
   pulls at $v=10$ \AA/ns) and L (63\%), with the distinction drawn at $t=0.3$
   ns. Inset: PMF profiles reconstructed considering the pulls of the
   two groups separately.  (b, c) The putative structural explanation
   of the water file disruption in group H. In these runs the water
   molecule preceding the pulled ion, W2, formed hydrogen bonds with
   the carbonyl atoms of the backbone close to the dimer interface.
   The bonds were most frequently observed with residues Val1 of chain
   A (about 50\% of the runs in group H) and Ala3 of chain B (70\%).
 }
 \label{fig:hbond-works}
\end{center}
\end{figure}

\begin{figure}
\begin{center}
 \includegraphics*[width=.7\columnwidth]{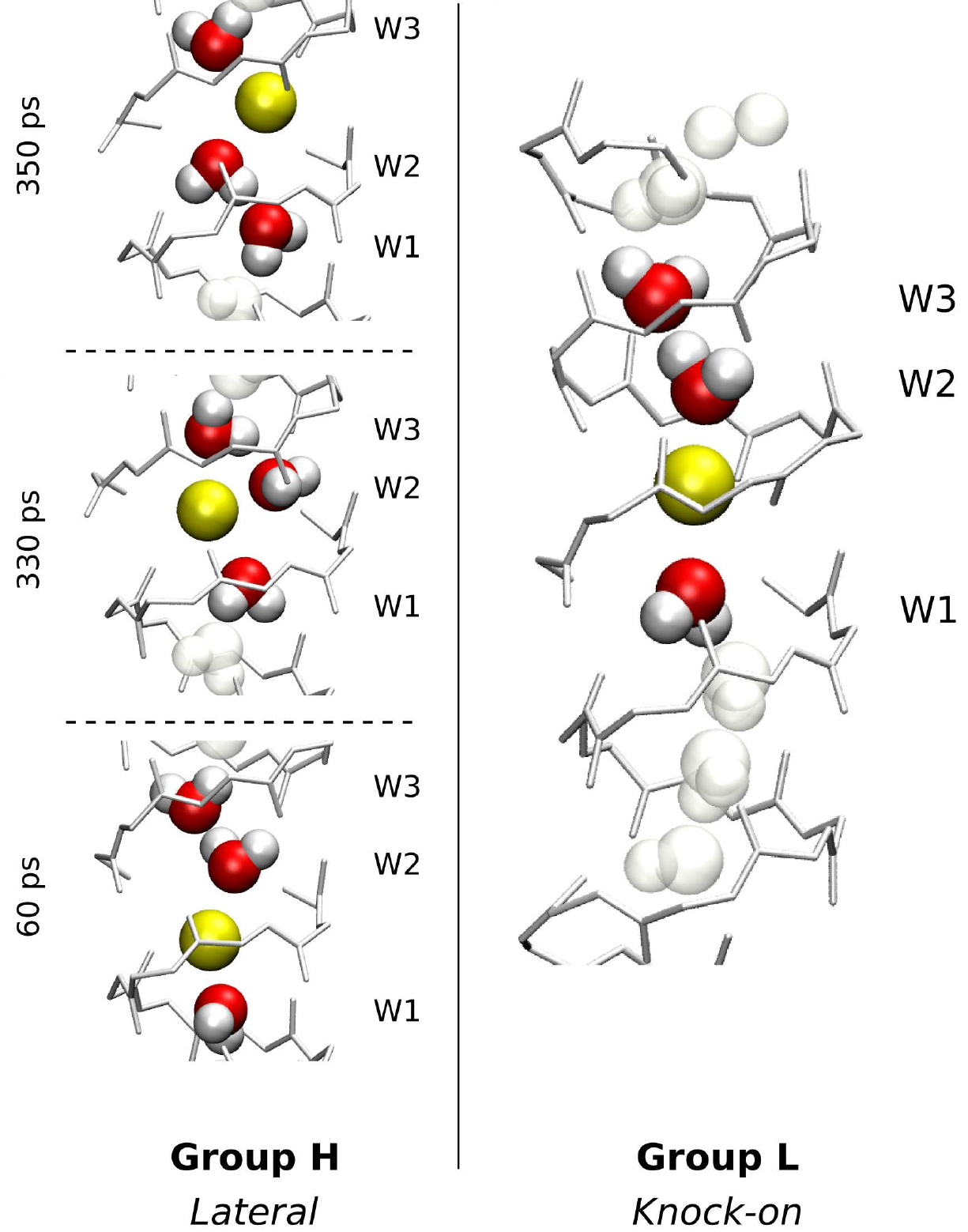}
 \caption{Left: the position of water molecules and $\ksmd$
   at three snapshots (60, 330 and 350~ps after the beginning of the
   run) during a representative trajectory of group H. The ion steered
   upwards (yellow) laterally ``slips past'' the preceding water
   molecule in the file (W2).  The water molecule is held in place by
   hydrogen bonds with carbonyl oxygens of the backbone and exchanges
   sides with the ion. Right: during runs in group~L the water
   filling the channel is displaced together with the ion, preserving
   the sequence of the file  (knock-on). }
\label{fig:HL}
\end{center}
\end{figure}

Ion and water permeate through the narrow gA channel as a single
file\cite{hille01,Allen2004}; the sequence of ion and water molecules should
therefore be preserved during a pull. However, during some of the
reverse (outgoing) pulls, $\ksmd$ was observed to exchange places with
the preceding water molecule. These trajectories could be
distinguished according to the value of the work $W$ cumulated at 300
ps, as shown in \ref{fig:hbond-works} (a).  We inspected the structural
features of the two groups analyzing a subset of the reverse runs,
and labelled the trajectories for which $W(300
\mbox{ ps}) > 10 $ kcal/mol as belonging to the ``H'' group (37\% of
the trajectories), and ``L'' otherwise (63\%). %
To analyze the atomistic basis for this difference,
we computed 52 additional trajectories
recording the state of the system every 10~ps, 
and labelled the pulls in groups H (14 pulls) and L (38
pulls) as above. Inspection of the trajectories in group H revealed
that the order in the water file was partially lost
around 300~ps after the beginning of
the pull (\ref{fig:hbond-works} (b) and (c)).  When this
 event occurred, the $\ksmd$ ion overtook the
preceding water molecule (indicated as W2 in \ref{fig:HL},
left).  Conversely, in the pulls in group L the order of the water
file is preserved (\ref{fig:HL}, right).

The interruption of the water file may be traced back to the formation
of hydrogen bonds between W2 and carbonyl atoms in the gA
backbone. \ref{fig:hbond-works} (b) and (c)  show the residues whose carbonyl atoms
were most often acceptors of a hydrogen atom of W2 at the time when it
was overtaken by $\ksmd$.  A hydrogen bond was observed with residue
Val1 of chain~A in about 50\% of the runs in group H, and with residue
Ala3 of chain B in about 70\% (chains~A and B being the monomer placed
at positive and negative $z$ respectively).  Radial and angular
cutoffs for H~bonds were taken as 4~\AA\ and $30^\circ$ respectively.

Finally, we performed a control simulation to check the
stability of the dimer's embedding in the hydrophobic membrane
environment.  Specifically, we checked the equilibrium configuration of the dimer
when the permeating ion was held close to the dimer interface ($z=0$
\AA) by a constant biasing potential of $k=10$ kcal/mol/\AA$^2$,
analogous to a setup that would be used for a US window in the middle
of the channel.  A simulation of 50 ns was sufficient to disrupt the
pore structure, with water fingering from bulk on the side of the
channel in order to balance the ion charge in the middle of the
membrane.  Thus, the biased equilibrium with the ion forced to stay
within the pore is substantially different from the unbiased
permeation event, which could imply that SMD is better than US
for this system.

\subsection{Sampling and convergence}

\begin{figure*}[tb]
\begin{center}
 \includegraphics[width=.9\columnwidth]{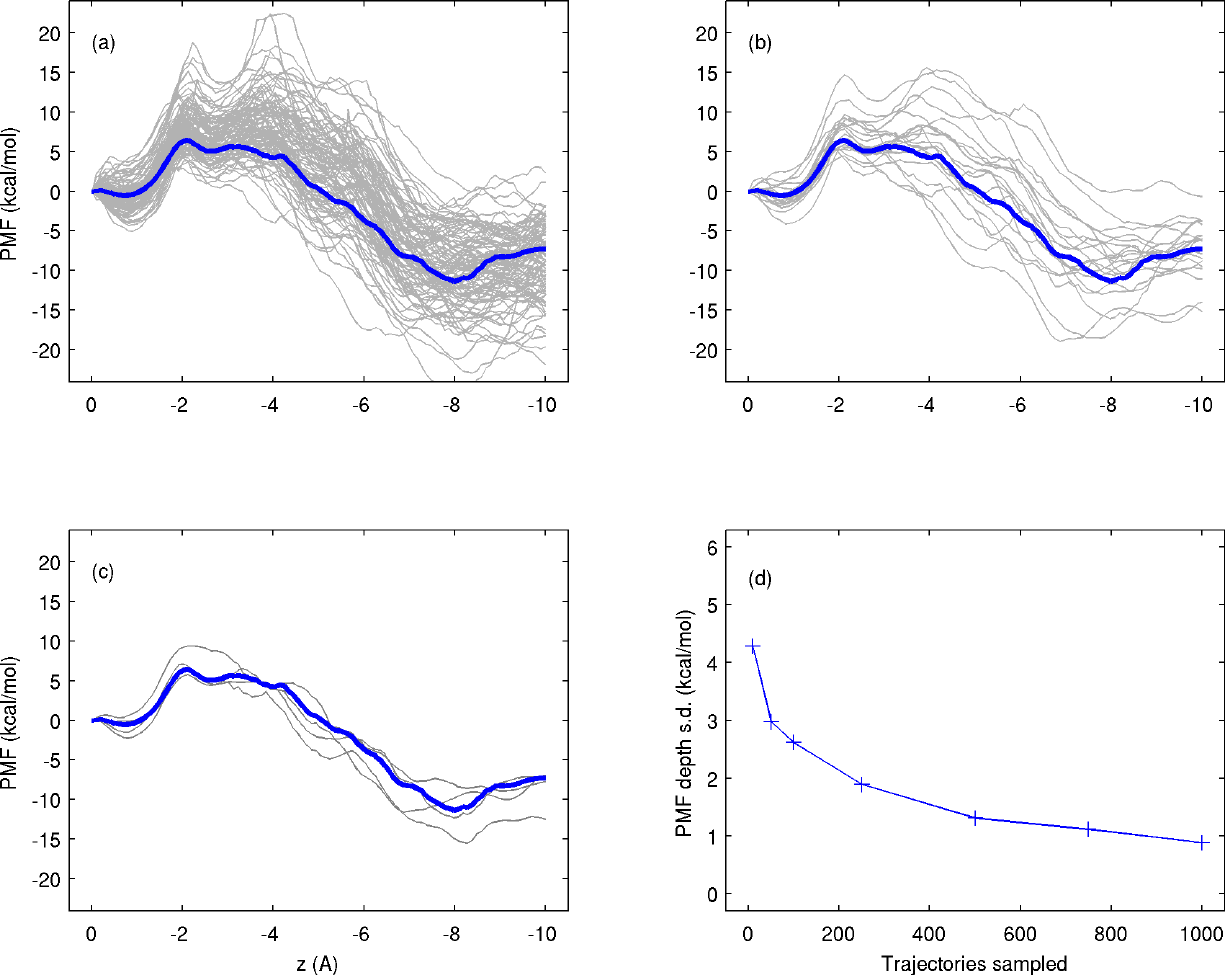}
\caption{ Potential of mean force curves computed in non-overlapping
  blocks of (a) 10 bidirectional pulls, (b) 50 pulls, and (c) 250
  pulls. Standard deviation of the PMF depth with various amounts of
  sampled trajectories, obtained by 200-fold bootstrapping (d).  }
\label{fig:confidence_bands}
\end{center}
\end{figure*}

We characterized the convergence of PMF estimates with respect to
increased sampling with two
methods. First, the available trajectories were split in
non-overlapping blocks of different sizes, computing the PMF profiles
using the data contained in each,  assuming
$G(0)=0$. \ref{fig:confidence_bands} (a) shows the PMF curves obtained
using 10 bidirectional trajectories each, i.e.  1-10 (first block),
11-20 (second block), and so on, for a total of 100 profiles.
Analogously, \ref{fig:confidence_bands} (b) shows PMF profiles computed
with blocks of 50 bidirectional pulls each (20 profiles), and \ref{fig:confidence_bands}
(c) shows the same using 250 bidirectional pulls at a time (4 profiles). For
reference, in panels (a)-(c) the thick blue line shows the PMF
computed with all available data.  Increased sampling clearly improves
the reproducibility of PMF curves; in particular, it appears that
curves obtained with only 10 bidirectional pulls each are affected by a
statistical error of the order of 10 kcal/mol, comparable 
to the PMF depth. 

We used the bootstrapping procedure outlined in Section \ref{sec:bootstr-proc}
to quantify the effect of increased sampling on the
variability of the PMF depth, $G(10)-G(0)$. \ref{fig:confidence_bands} (d)
shows the standard deviation of the PMF depth over the bootstrapped
profiles, obtained including different number of trajectories. The
results confirm that the standard deviation for $R=10$ trajectories is
of the order of 5 kcal/mol, which decreases to 3 for $R=50$ and to 2
for $R=250$. When using all of the 1,000 available trajectories, the
bootstrap analysis estimates a statistical error of 1~kcal/mol.

Finally, we performed two additional simulation sets as controls to
check the influence of pulling speed on the PMF depths and
profiles. The structures for these runs were taken from the previous
study~\cite{DeFabritiis2008} and were slightly smaller than the
production simulations ($\sim$ 29,000 atoms in total). In the two control
sets the ion was pulled at $v=10$ (221 bidirectional pulls) and
$v=2.5$ \AA/ns (171 bi-pulls) respectively. We used the
afore-mentioned bootstrapping technique to compute the convergence of
the PMF depth in the two data sets (\ref{tab:speed-pulls}).  The final
PMF profiles are qualitatively similar with each other and with the
one obtained from the production simulations. Consistently to what is
observed in the production runs, decreasing the statistical error
below 2 kcal/mol requires significant computational effort for both
pulling speeds. Given that slow pulls require four times as much
simulation time as the fast ones, performing SMD at $v=10$ \AA/ns
appears to be more computationally efficient for this system.

\begin{table}
  \caption{Statistical uncertainty of the PMF depth estimate (in
    kcal/mol), at ion pulling speeds of $v=10$ and $v=2.5$ \AA/ns,
    computed through bootstrapping for increasing number of pulls. The
    figures show the width of the 68\% confidence interval (CI), matching a
    1-$\sigma$ interval around the mean. Computation of each of the
    slower pulls requires four times the computational effort of a
    fast one.}
  \label{tab:speed-pulls}
  \begin{tabular}{ccc}
\hline \hline
\multicolumn{3}{c}{\bf PMF depth CI (kcal/mol)} \\
Pulls &	10 \AA/ns & 2.5 \AA/ns  \\
\hline 
10    &	8.3    & 8.7	  \\
25    &	5.4    & 6.9	  \\
50    &	3.9    & 5.8	  \\
100   &	3.2    & 4.3	  \\
150   &	2.8    & 3.9	  \\
171   &	--     & 3.6	  \\
221   &	2.5    & --	  \\
\hline \hline
  \end{tabular}
\end{table}

\section{Discussion and conclusions}

In this
paper we performed an extensive set of SMD forward-reverse
experiments, sampling a simple but realistic biomolecular test system, gA, well
beyond the state of the art, to analyze the effect of increased
sampling on the precision of the estimate of the PMF of permeation of
a potassium ion steered through the channel.

The advent of accelerator processors and codes able to exploit
them, like ACEMD for graphical processing units~\cite{Harvey2009}, should play an
important role in reducing sampling limitations to obtain  free
energy estimates  up to the accuracy of the force-fields. 
The SMD protocol is well suited for high-throughput molecular
simulations in distributed computing infrastructures like GPUGRID.
Still, considering the particle being steered, the case of ion permeation through
gA is a relatively simplistic test case.  In particular, a
single ion does not have orientational nor conformational degrees of
freedom; this is not the case  when dealing with generic
ligands, whose internal degrees of freedom have to be sampled, as well. 
The scientist has to prepare two initial systems (forward and reverse)
from which many simulations are spawned.  
Umbrella sampling offers a similar degree of parallelism, as long as
several uncorrelated configurations are generated to start each
US window or a sufficient equilibration time (hundred nanoseconds or
more) is allowed for each window~\cite{buch_high-throughput_2010}.  US
simulations with stratification of windows showed much less variability
than the one reported here for SMD, but the question remains of how much is
actually due to poorer sampling.  As each US window is starting from a
single initial configuration, the US protocol is averaging the
effective potential on a small area of the configurational space close
to it. If multiple configurations are used, as in
Ref.~\cite{buch_high-throughput_2010}, similar levels of fluctuations
in the free energy are obtained.  With US, the method used
to generate the initial configurations of the biasing windows has a
crucial importance but it is subject to the specific choice of the
scientist; the SMD protocol mitigates the problem of
generating initial conditions because better sampling is achieved
simply by increasing number of trajectories.
However, for gA, the US biased equilibrium with the ion forced to
stay within the pore was structurally different from the unbiased configuration,
and thus probably less representative of a
permeation event than a SMD pull.

Finally, the choice of the pulling speed  influences the amount of
pulls for a fixed computational cost:  pulling too fast
would produce higher-energy pathways but allow for more pulls, while 
slower pulls would be closer to equilibrium but more computationally
demanding.  Given that the
system is in non-equilibrium, however, there may be regions 
of the phase space that only
become accessible after some transient time, like for the two permeation
pathways showed for the potassium ion in gA, and faster
 pulling speeds may prevent some conformational transitions from
happening.  Therefore, the preference of SMD
versus US is probably system-dependent, and the biasing methodology
that  produces the lesser perturbation compared to the
unbiased case should be chosen.

In the case studied here, thousands of pulls were required to reach a
statistical precision within one kcal/mol.  Even though the results
were obtained on the basis of extensive experiments on a specific,
admittedly simple, system, similar considerations may apply to more
complex cases. %

\section{Acknowledgments}

The authors would like to thank Dr.\ S.\ K.\ Sadiq, I.\ Buch, and
M.\ J.\ Harvey for reading the manuscript. We acknowledge Sony
Computer Entertainment Spain. 
GDF acknowledges support from the Ramon
y Cajal scheme and by the Spanish Ministry of Science and Innovation (Ref. FIS2008-01040).
TG acknowledges support from the Programa Beatriu de Pin\'{o}s from the Generalitat de
Catalunya.  The authors would like to express a special
acknowledgment to the contribution of the volunteers who participate
to GPUGRID (formerly PS3GRID) distributed computing networks. This
work was partially supported by the virtual physiological human EU
network of excellence (VPH-NoE).

\vfill

\ifx\mcitethebibliography\mciteundefinedmacro
\PackageError{achemso.bst}{mciteplus.sty has not been loaded}
{This bibstyle requires the use of the mciteplus package.}\fi

\end{document}